\documentclass{elsart}

\usepackage{natbib}

\usepackage{graphicx}
 
\begin{document}


\runauthor{Snellen et al.}


\begin{frontmatter} 
\title{Faint Gigahertz Peaked Spectrum sources and the evolution of 
young radio sources} 
\author[IOA]{I.A.G. Snellen}
\author[JIVE,LEIDEN]{R.T. Schilizzi}
\author[LEIDEN]{G.K. Miley}
\author[Bristol,IAP,LEIDEN]{M.N. Bremer}
\author[LEIDEN]{H.J.A. R\"ottgering}
\author[JIVE]{H.J. van Langevelde}

\address[IOA]{Institute of Astronomy, Madingley Road, Cambridge, U.K.}
\address[JIVE]{Joint Institute for VLBI in Europe, Postbus 2, Dwingeloo, The Netherlands} 
\address[LEIDEN]{Leiden Observatory, Postbus 9513, Leiden, The Netherlands}
\address[Bristol]{Department of Physics, Bristol University, Tyndall Avenue,
Bristol, U.K.}
\address[IAP]{Institut d'Astrophysique de Paris, 98bis Boulevard Arago, Paris, France}

\begin{abstract} 

GPS sources are the objects of 
choice to study the initial evolution of extragalactic radio sources, since 
it is most likely that they are the young 
counterparts of large scale radio sources.
Correlations found between their peak frequency, peak flux density and 
angular size provide strong evidence that synchrotron self absorption
is the cause of the spectral turnovers, and indicate that young radio sources
evolve in a self-similar way. The difference in redshift 
distribution between young and old radio sources must be due to a 
difference in slope of their luminosity functions, and we argue that this 
slope
is strongly affected by the luminosity evolution of the individual sources.
A luminosity evolution scenario is proposed in which GPS sources
increase in luminosity and large scale radio sources decrease
in luminosity with time. It is shown that such a scenario agrees with 
the local luminosity function of GPS galaxies.

\end{abstract} 


\begin{keyword}

radio continuum: galaxies

\end{keyword}

\end{frontmatter}

\section{GPS versus CSO}

Gigahertz Peaked Spectrum (GPS) sources optically identified with
galaxies are most likely to possess compact symmetric radio
morphologies \citep{St1}. In addition, the large majority of 
Compact Symmetric Objects (CSO) exhibit a gigahertz-peaked 
spectrum. The large but not complete overlap between these two classes
of sources is caused by the synchrotron
self-absorbed mini-lobes, located at the extremities of most CSOs, being the
main contributors to the overall radio spectrum, and producing the peak at 
about 1 GHz in frequency. 
However, not all CSOs are GPS galaxies. If a CSO
is oriented at a small angle to the line of sight, the contributions of
the central core and fast-moving jet feeding the approaching mini-lobe
become important, producing a flatish and variable spectrum 
\citep{S5}. 
Observed at such a small viewing angle,  the large contrast between 
the approaching and receding side of the radio source makes it also 
increasingly difficult to identify the object as a CSO, and explains why only 
a small fraction of CSOs is identified with quasars.
GPS sources identified with quasars have (almost) similar radio spectral 
properties to GPS galaxies.
However, it is unlikely that these high redshift objects are unified
with GPS galaxies by orientation, and are probably unrelated \citep{S3,S4}.

CSO and GPS galaxies have recently been proposed \citep{F2,R2} as the young 
counterparts 
of ``old'' FRI/FRII extended objects. The alternative hypothesis that GPS 
sources are themselves old and situated in particularly dense media seems 
unlikely, since 
their surrounding media are insufficiently dense to confine them \citep{F2}. 
Conclusive evidence that they are young has been given by Owsianik and 
collaborators
\citep{Ow1} who measured the propagation velocity of the hot spots of 
several prototype GPS sources and CSOs to be $\sim 0.2$c, which implies 
dynamical ages of typically $\sim 10^3$ years. 

Although research has always been concentrated 
on the question of whether GPS sources are young or old and situated in a 
dense medium, 
the recent breakthrough on this matter does not mean that the subject
of GPS and CSOs is now closed. On the contrary, GPS sources and 
CSOs are the objects of choice to study the initial evolution of extragalactic
radio sources.
  
\section{A sample of faint GPS sources}

It is relatively straightforward to select young radio sources on the 
basis of their GHz peaked spectrum. The disadvantage of this approach is
that some sources with CSO morphology will be missed. However, for the very 
compact and/or faint radio sources in particular, it is virtually impossible 
with the current 
resolution and sensitivity of VLBI networks to establish 
whether they are CSOs, while it remains possible to determine their
radio spectra. 

Here we present results on a sample of faint GPS sources selected from
the Westerbork Northern Sky Survey \citep{R1} using
additional observations with the WSRT and the VLA \citep{S6}.
It consists of 47 sources with peak frequencies between 500 MHz and 
15 GHz, and peak flux densities ranging from $\sim 30$ to $\sim 900$ mJy. 
The combination of this new faint sample and existing brighter samples
\citep{F1,St2} allow for the first time
the disentanglement of redshift and radio luminosity effects.
The sample has been studied extensively in the optical to determine
the nature and redshifts of the optical identifications, resulting in 
an identification fraction of $\sim 87\%$ \citep{S3}, and
redshifts for $\sim 19\%$ of the sources \citep{S4}.
About 40\% of the sample consists of high redshift quasars (which we will 
further ignore), a fraction comparable to that found in bright GPS samples.
Only a few of the redshifts of GPS galaxies have been determined yet,
due to their faint magnitudes and weak emission lines. Fortunately their
redshifts can be estimated due to their well established Hubble diagram
\citep{S1}. 
Global VLBI observations at 5 GHz were obtained for all sources in the 
sample. In addition, VLBA observations at 
15 GHz were obtained for the sources with peak frequencies $\nu_p>5$ GHz, 
and global VLBI observations at 1.6 GHz for sources with $\nu_p<5$ GHz which
were found to be extended at 5 GHz (Snellen et al., in prep.).  
The angular sizes of the GPS galaxies, measured as the distance between
the two outermost components or the deconvolved FWHM for
a single component source, were found to be between 0.5 and 33 mas, while
the sizes of the GPS galaxies in the bright \citet{St2} sample 
range from 6 to 350 mas.

\section{Morphological evolution of young radio sources}

In addition to the correlation between peak frequency $\nu_p$ and 
angular size $\theta$ as found in the bright samples (the higher $\nu_p$,
the smaller $\theta$, eg. Fanti et al. 1990), we found a correlation
between the peak flux density $S_p$ and $\theta$ (the higher $S_p$, the 
larger $\theta$). The strength of these correlations are as expected from 
Synchrotron Self Absorption (SSA) theory, in which $\theta^2 \propto S_p 
\nu_p^{-5/2}$,
and therefore provide strong evidence that the spectral peak is indeed 
caused by SSA and not by free-free absorption, as proposed by 
\citet{B1}. 
The spectral peak mainly originates in the dominant features of the 
radio source, the mini-lobes, and 
therefore reflects the sizes of the mini-lobes. Note that $\theta$ is the 
overall size of the radio source, eg. the distance between the two mini-lobes
for the correlations discussed above. The correlations
between $\nu_p$, $S_p$ and $\theta$ imply a linear correlation between
the mini-lobes and overall sizes, meaning that during the evolution of young 
radio sources the ratio of the size of the mini-lobes and the distance between
 the two minilobes is constant. This suggests they evolve in a 
self-similar way \citep{S2}.
\begin{figure}[hbt]
\centering
\includegraphics[scale=0.4,angle=-90]{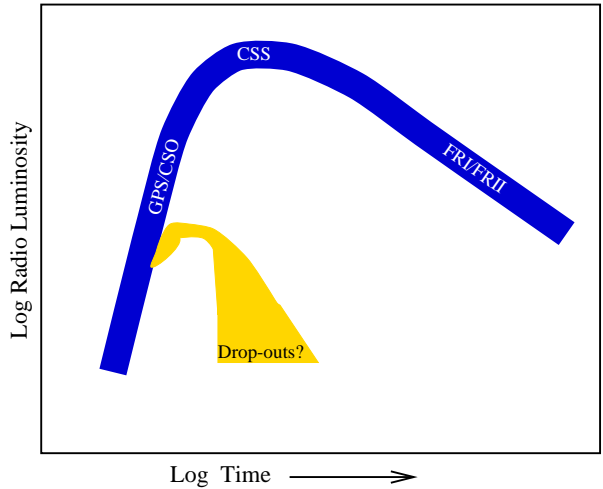}  
\caption{The luminosity evolution proposed for extragalactic 
radio sources.}
\end{figure}

\section{Luminosity evolution of young radio sources}

The lifetimes of radio sources are short compared to cosmological
timescales. The populations of young and old radio sources should therefore
undergo identical cosmological evolution. However, in flux density limited
samples, GPS galaxies are found at higher redshifts than large scale 
radio galaxies (e.g. Snellen 1997). This can only mean that the ratio
between low and high luminosity sources is smaller (eg. a flatter
luminosity function) for the GPS galaxies than for large scale radio galaxies. 
We argue that the slope of the luminosity function is strongly dependent
on the luminosity evolution of the individual sources, and propose an 
evolution scenario in which  radio sources in their GPS phase increase
in luminosity with time 
and decrease in luminosity when they become large scale radio sources
(figure 1).
Ignoring source to source variations in the surrounding media,
their luminosities are only dependent on age and jet power.
Sources in a volume-based sample are biased towards low jet-powers and older 
ages, for populations of both GPS and large scale sources. 
Low jet powers result in low luminosity sources. The higher the age 
of a large scale source the lower its luminosity, but the higher the age of
a GPS source the higher its luminosity. This means that for a population of 
large scale sources the jet power and
age biases strengthen each other resulting in a steep luminosity function,
while they counteract for GPS sources, resulting in a flatter luminosity 
function. 
The luminosity evolution proposed is expected for a ram-pressure confined
radio source in a surrounding medium with a King profile density. 
In the inner parts of the King profile, the density of the medium is constant 
and the radio sources builds up its luminosity, but after it grows 
large enough the density of the surrounding medium declines and the 
luminosity of the radio source
decreases. A very simple model of such kind was presented by Snellen (1997),
but much more sophisticated models have been constructed by 
\citet{K1}.

Comparison of the local luminosity function (LLF) of young and old radio 
sources can put strong constraints on the rise and decay of the 
radio luminosity, the age ratio between the old and young sources, and
the percentage of `drop-outs', young sources which are short lived and 
will never end up as large scale radio sources (figure 1). 
Unfortunately, because GPS sources are biased towards high redshifts, only
a handful of young sources are known at $z<0.2$, insufficient to construct
a LLF. However, the cosmological evolution of the radio luminosity function
of large scale radio sources is more or less known \citep{D1},
and we can assume that young radio sources undergo an identical cosmological
evolution. In this way, all the sources in the faint and bright GPS samples
can be used to construct a LLF of young radio sources. 
The exact redshift of each source is not important, 
because the `luminosity-evolution' of the radio luminosity function
counteracts the redshift dimming; Sources with a similar flux density,
located between $0.5<z<1.5$, contribute to the same luminosity bin. 
The combination of the faint and bright samples had to be done with 
great care, since they are selected in different ways (on the 
optically thick 
and thin part of their spectrum respectively), and therefore different 
selection effects 
play a role (Snellen et al., in prep.)
The resulting LLF is shown in figure 2. 
\begin{figure}[hbt]
\centering
\includegraphics[scale=0.5]{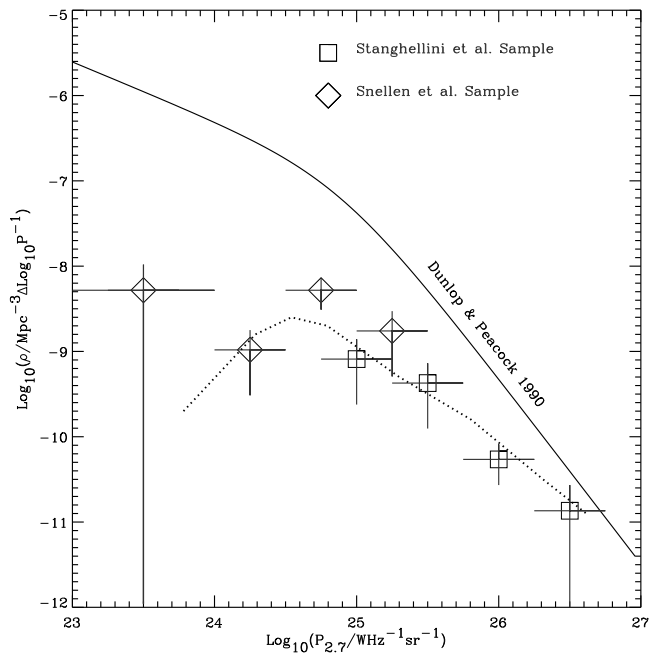}  
\caption{The Local Luminosity Function (LLF) for GPS galaxies. 
The dotted line indicates the LLF for GPS galaxies, as expected for 
a population of radio sources undergoing the rise and decay in luminosity
as described in this paper.}
\end{figure}
The dotted line indicates a simulation of the LLF for GPS galaxies,
assuming that a population of radio sources undergoes the rise and decay in 
luminosity
as described above, and has a power law distribution of jet powers. 
Their luminosities and number counts were scaled in such way that the LLF of 
large radio sources matches the LLF of FRI/FRII as presented by \citet{D1}.
Although the uncertainties on the datapoints are large and several free
parameters enter the simulation, it shows that the shape of the LLF of GPS 
sources is as expected, and that large and homogeneously
defined samples of GPS sources can constrain the luminosity
evolution of extragalactic radio sources.

{\it This research was in part funded by the European Commission under
contracts  ERBFMRX-CT96-0034 (CERES) and 
ERBFMRX-CT96-086 (Formation and Evolution of Galaxies), and
SCI*-CT91-0718 (The Most Distant Galaxies).}

\end{document}